\documentclass[%
 reprint,
superscriptaddress,
 amsmath,amssymb,
 aps,
]{revtex4-2}

\usepackage{graphicx}
\usepackage{dcolumn}
\usepackage{bm}
\usepackage{soul}

\sethlcolor{yellow}

\usepackage{color}
\usepackage{float}
\usepackage{graphicx}
\usepackage{subcaption}
\usepackage{booktabs}
\usepackage{graphicx}
\usepackage{subcaption}

\usepackage[colorlinks=true, urlcolor=blue, linkcolor=red, citecolor=green]{hyperref}

\usepackage[normalem]{ulem}
\begin{document}

\preprint{APS/123-QED}

\title{Precision Ringdown Measurements of Binary Black Hole Remnants}
\author{Achal Kumar}
\affiliation{Physics Department, University of Florida, PO Box 118440, Gainesville, FL 32611-8440, USA}

\author{Poulami Dutta Roy}
\affiliation{Physics Department, University of Florida, PO Box 118440, Gainesville, FL 32611-8440, USA}

 \author{Marek~J.~Szczepa\'nczyk}
\affiliation{Faculty of Physics, University of Warsaw, Ludwika Pasteura 5, 02-093 Warsaw, Poland}

\author{Sergey Klimenko}
\affiliation{Physics Department, University of Florida, PO Box 118440, Gainesville, FL 32611-8440, USA}




\date{\today}

\begin{abstract}
The ringdown gravitational wave from a binary black hole (BBH) merger is a superposition of quasi-normal modes (QNMs) of the remnant black hole. In general relativity (GR), QNMs are damped harmonic oscillations with frequencies and damping times uniquely determined by the remnant's mass and spin. The measurement of the ringdown modes and performing black hole spectroscopy provides a tool to test the validity of GR.
In this work, we present \texttt{RingCWB}, a ringdown analysis method based on coherent WaveBurst (cWB), an unmodeled pipeline for the detection and reconstruction of gravitational-wave signals.
This method yields tighter constraints on the QNM frequency and damping time than previous measurements. The improved precision results from the noise reduction achieved by the cWB reconstruction and the enhanced ringdown analysis, which probes the remnant properties at earlier times, closer to the merger. We have analyzed publicly available binary black hole (BBH) detections from the third Gravitational-Wave Transient Catalog (GWTC-3). For all events considered, the measured frequency and damping time of the dominant $(l,m)=(2,2)$ mode are found to be consistent with the predictions of GR. A combined analysis further strengthens these constraints, yielding fractional deviations in frequency $\delta f_{220} = 
-0.005_{-0.028}^{+0.028}$ and damping time $\delta\tau_{220} =
0.032_{-0.090}^{+0.108}$, consistent with zero within the quoted uncertainties.
\end{abstract}

\maketitle



\section{Introduction}

Over the past decade, the Laser Interferometer Gravitational-Wave Observatory (LIGO) has enabled detection of gravitational wave (GW) signals originating from compact binary coalescences. These observations are consistent with mergers of binary black holes (BBH) \cite{LIGOScientific:2016aoc}, binary neutron stars (BNS) \cite{LIGOScientific:2017vwq}, and neutron star–black hole (NSBH) systems \cite{LIGOScientific:2021qlt}. They enable the study of gravity in the strong-field, high-velocity and  dynamical regime, thereby providing new tests of Einstein’s general relativity (GR). These tests will complement well established tests of GR, including Solar System tests, binary-pulsar experiments, observations of massive black holes (BH) at galactic centers, and cosmological measurements \cite{Will:2014kxa,Berti:2015itd,Freire:2012mg,GRAVITY:2018ofz,Do:2019txf,EventHorizonTelescope:2019dse,Clifton:2011jh}, which probe gravity in low-velocity, quasi-static and weak-field regimes. Although GR has successfully passed all experimental tests to date,  it still has open questions associated with BH, such as their stability 
\cite{Dafermos:2016uzj,TeixeiradaCosta:2019skg,Klainerman:2021qzy,Dafermos:2021cbw}, the existence of singularities inside their event horizon \cite{Penrose:1964wq,Hawking:1970zqf}, and Hawking’s information-loss paradox \cite{Hawking:1976ra,Almheiri:2020cfm,Raju:2020smc}. Moreover, GR is incomplete in the quantum regime and requires the introduction of dark matter and dark energy to account for cosmological observations. These limitations motivate the search for possible deviations from GR and for extensions of the theory.  

The gravitational wave (GW) signal emitted by a BBH system can be divided into three phases. During the inspiral phase, black holes steadily approach each other, losing energy and angular momentum through GW emission. As they come sufficiently close, the system enters the merger phase - a highly dynamical regime in which the black holes coalesce, forming a highly perturbed remnant. In the final ringdown phase, the perturbed remnant radiates GWs as it stabilises into a Kerr BH.

Once a GW signal is identified by search algorithms \cite{KAGRA:2021vkt,LIGOScientific:2025hdt}, the data from the GW detector network is used to infer the properties of the source. The full inspiral-merger-ringdown (IMR) signal is matched to simulated GR waveforms to obtain posterior probability for the source parameters \cite{Cutler:1994ys}, typically assuming Gaussian, stationary noise that is uncorrelated between detectors \cite{LIGOScientific:2019hgc,Veitch:2014wba,Christensen:2022bxb,LIGOScientific:2025yae}. 
The best-fitting IMR waveform obtained from the parameter estimation represents the signal expected in GR.  The inferred source parameters can then be used for a range of further analyses, including population studies~\cite{KAGRA:2021duu,LIGOScientific:2025pvj}, measuring cosmic expansion history~\cite{LIGOScientific:2025jau} and tests of GR \cite{LIGOScientific:2025rid,LIGOScientific:2021sio}.

Among the stages of a GW signal, the ringdown provides one of the cleanest tests of GR. It is governed by the perturbative dynamics of the remnant black hole, which is well described within GR~\cite{Berti:2025hly,Chandrasekhar:1975zza}. A linearly perturbed Kerr BH is expected to emit GW radiation described by quasinormal modes (QNMs), which are damped harmonic signals characterised by specific frequencies and damping times~\cite{Vishveshwara:1970zz,Berti:2009kk}. In GR, the QNM parameters depend solely on the mass and spin of the remnant BH and are independent of the properties of the progenitor binary or the details of the merger. The QNMs represent the characteristic BH spectrum that uniquely identifies it as described by GR. 

A precise measurement of this spectrum would provide a stringent test of GR and a direct insight into the nature of BHs. However, such tests typically require the detection of multiple ringdown modes. This is challenging because the ringdown signal is relatively weak, and the subdominant modes, beyond the dominant (2,2) mode, have significantly smaller amplitudes, making them much harder to detect. Therefore, current ringdown analyses adopt alternative strategies for testing GR, rather than attempting the direct BH spectroscopy.
For example, the frequency-domain analysis, based on the \texttt{SEOBNRv5PHM} waveform model~\cite{Brito:2018rfr, Ghosh:2021mrv,Maggio:2023vch}, introduces fractional deviations in the frequency and damping time of the $(2,2)$ QNM within the ringdown description of the model. These deviations are then constrained by using the entire GW signal. Another class of ringdown analysis~\cite{Carullo:2019flw} isolates the ringdown by selecting a data segment of the post-merger signal dominated by QNMs. The segment start time is defined relative to the peak time of the best-fit IMR waveform. Within the selected data segment, the ringdown is modelled as a superposition of damped sinusoids, directly inferring either the QNM frequencies and damping times or the remnant BH’s mass and spin. The ringdown parameters are then compared with the values 
inferred from the IMR parameter estimation with GR waveforms.

These approaches, however, rely on different assumptions that can influence the inferred ringdown parameters. In particular, the methods that incorporate deviations directly within full waveform models are inherently model dependent. Therefore, the ringdown parameter estimates are affected by systematic uncertainties associated with the underlying waveform model. 
Methods that directly fit QNMs typically select the post-merger data segment well after the peak amplitude, where the signal-to-noise ratio (SNR) has decreased substantially. 
This significantly reduces the precision of the measured ringdown parameters. In addition, these analyses are susceptible to non-Gaussian noise fluctuations in the detector data, whose impact on ringdown parameter estimation is difficult to model. 

In this paper, we present \texttt{RingCWB}, a ringdown analysis that addresses these limitations. Specifically, 
(i) the ringdown analysis is performed on the signal recovered with cWB instead of the raw strain data. cWB is a model-agnostic search algorithm which exploits coherence across multiple detectors to identify and extract denoised GW signals~\cite{Mishra:2024zzs,Klimenko:2005xv,Klimenko:2008fu,Klimenko:2015ypf}. Therefore, the analysis of the cWB-reconstructed signals is less susceptible to systematic errors due to detector noise.
(ii) A robust and model-independent method for isolating ringdown from the full GW signal is introduced. This allows the analysis to start closer to the merger, yielding tighter constraints on the ringdown parameters. (iii) Finally,  a dedicated study is carried out to assess potential reconstruction biases and systematic uncertainties, as well as to study the impact of non-Gaussian noise on the ringdown measurements. Selected GW events from the GWTC-3~\cite{KAGRA:2021vkt} were analysed. 


\begin{figure*}[t]
    \centering

    \includegraphics[width=\textwidth]{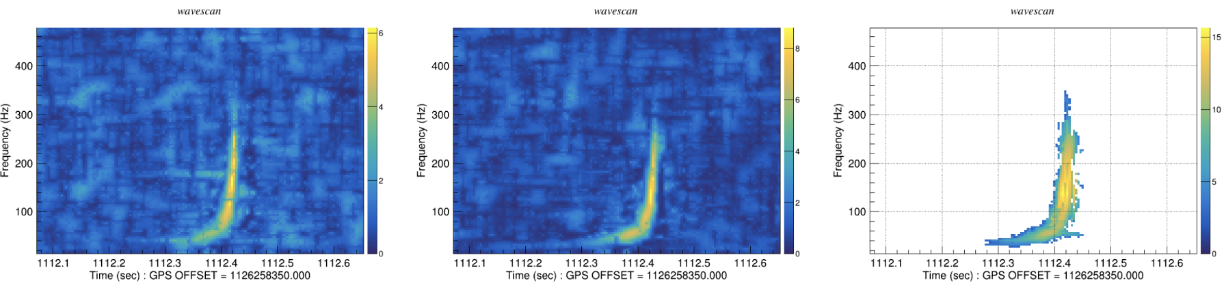}


    \caption{\label{fig:ced_150914} TF representation of the first event GW150914 detected by LIGO: left panel - Livingston detector, middle panel - Hanford detector, right panel - selected cross-power pixels. }
\end{figure*}

\section{ Method}
\label{sec:Method}

\subsection{cWB reconstruction}
\label{sec:cwb}
Coherent WaveBurst (cWB) is an unmodeled GW search algorithm designed to identify transient GW signals in the data from a network of detectors~\cite{Mishra:2024zzs}. Unlike matched-filter searches, cWB does not rely on a specific waveform model. Instead, it searches for coherent signal power in multiresolution time-frequency (TF) representations of the detector strain data~\cite{Klimenko:2022nji}. The analysis uses strain data from the two LIGO detectors, Livingston and Hanford. As an illustration, Figure~\ref{fig:ced_150914} shows the TF representation of the first detected GW event GW150914~\cite{LIGOScientific:2016aoc}  obtained with the WaveScan transform~\cite{Klimenko:2022nji}. Figure~\ref{fig:ced_150914} (right) highlights the distribution of coherent signal power across selected TF data samples, or pixels.  Once the coherent events are identified, cWB reconstructs the source sky location and the GW signal amplitudes in the TF domain using a constrained maximum-likelihood method~\cite{Klimenko:2005xv}. The corresponding time-domain waveforms (Figure~\ref{fig:rec_eng}, left panel) are obtained by applying the inverse WaveScan transform to the reconstructed TF amplitudes. 

In this work, the reconstructed cWB waveforms, rather than the raw strain data, are used for the ringdown parameter estimation. In this context, cWB acts as a denoising algorithm that mitigates the impact of detector noise on the ringdown reconstruction. The method substantially reduces reconstruction biases arising from non-Gaussian and non-stationary noise artefacts while suppressing Gaussian noise fluctuations by approximately a factor of $\sqrt{2}$. The primary drawback is the introduction of a systematic bias in the reconstructed ringdown frequency associated with the incomplete recovery of the signal power, particularly for weak signal components with low SNR. As discussed in Section~\ref{subsec:rec_bias}, however, this bias is not large and can be accounted for with the SNR-dependent correction.



\subsection{Fitting Procedure}
\subsubsection{Ringdown Model}
\label{subsec:ring_model}

Following the predictions of GR, the ringdown signal is modelled as a sum of damped sinusoids, corresponding to different QNMs, 
parametrised by the remnant's mass and spin.
Respectively, the Kerr model is used to relate mass and spin to the QNMs frequencies and damping time. 
Since the subdominant QNMs are  weak, 
the current analysis uses a two-mode model with the dominant $(l,m,n) = (2,2,0)$ mode and its overtone $(l,m,n) = (2,2,1)$. 

\begin{equation}
\begin{split}
h(t) = Ae^{i[2\pi f_{0} + \phi_{0}]} + \epsilon Ae^{i[2\pi f_{1} + \phi_{1}]} ,
\end{split}
\end{equation}

where $A$ denotes the amplitude of the $(2,2,0)$ mode. The parameter $\epsilon$ represents the relative amplitude, such that $\epsilon A$ is the amplitude of the $(2,2,1)$ mode. The parameters $(f_0,\tau_0)$ and $(f_1,\tau_1)$ are the frequency and damping time of the $(2,2,0)$ and $(2,2,1)$ modes, respectively, with $\phi_0$ and $\phi_1$ being the corresponding phases. Since the frequencies and damping times are functions of the detector frame remnant mass $(M_f)$ and spin $(a)$, we have a six-dimensional parameter space $\{A,M_f,a,\epsilon,\phi_1,\phi_2\}$. Bayesian inference with BILBY~\cite{Ashton:2018jfp,Romero-Shaw:2020owr} is performed to obtain the posterior distribution of the parameters. 


\subsubsection{Ringdown Window}
\label{subsec:ring_win}
The ringdown window defines a segment of the post-merger data used for the ringdown analysis where the nonlinear merger effects are small. Identifying this window requires a reference time in the evolution of the GW signal.
One widely adopted method is to use the signal peak amplitude as the reference point. The window is selected a certain time away from the peak time to ensure that analysis is in the ringdown phase. But this approach has the following drawbacks: (i) the peak amplitude is sensitive to several effects (like whitening artifacts, higher-order modes, and non-Gaussian noise) and, therefore, not a robust reference point.(ii) Typically, the ringdown analysis begins at 8–10 $t_{M_f}$ after the peak time, where $t_{M_f} = GM_f/c^3 $ is the characteristic timescale defined by the remnant mass $M_f$ in the detector frame. However, the signal SNR decreases substantially by then, leading to less precise measurements. 
\begin{figure}[htbp]
    \centering
    \hspace{-0.8cm}  \includegraphics[width=0.51\textwidth]{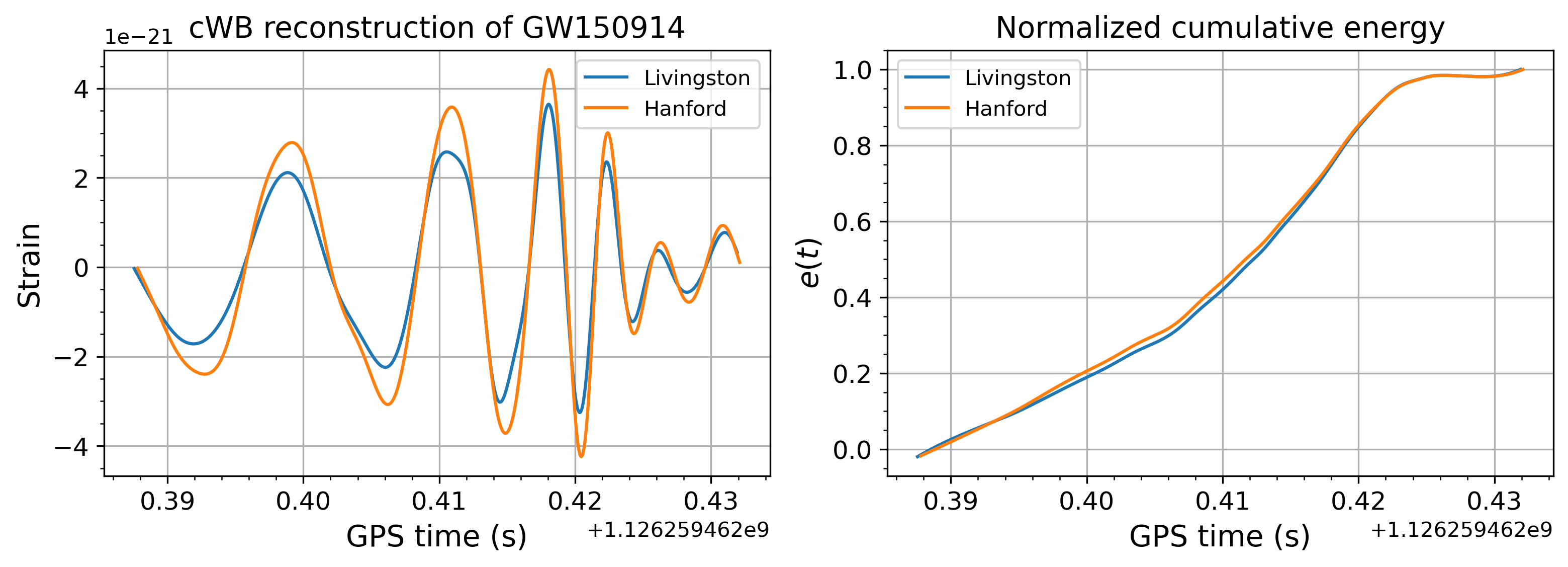}
    \caption{\label{fig:rec_eng} The reconstructed time-domain signals (left panel) and the normalized cumulative energy (right panel) as a function of time for GW150914.}
\end{figure}


To address these issues, the reconstructed signal is used to calculate the cumulative energy $E(t)$ defined as follows:
\begin{equation}
    \label{E_c_def}
    E(t) = \int_{ T_{\text{start}}}^{t}(h^2(\tau) + H^2(\tau))d\tau,
\end{equation}
where, $h(t)$ and its $90^\circ$ phase-shifted quadrature $H(t)$ are  the reconstructed GW waveforms. The integration limits, $(T_{\text{start}}$ and $T_{\text{end}})$, are set to three cycles before and four cycles after the peak of the waveform envelope, respectively. This choice ensures that the shape of the normalized cumulative energy curve
\begin{equation}
 e(t)=E(t)/E(T_{\text{end}})
 \end{equation}
 remains nearly independent of the remnant mass over a broad mass range.   
For example, Fig \ref{fig:rec_eng} shows the cWB reconstructed signal (left panel) and the normalized cumulative energy curve $e(t)$ (right panel) for the event GW150914~\cite{LIGOScientific:2016aoc}.

The start time $T_\text{w}$ of the ringdown window is defined as the time where $e(T_\text{w})=0.82$. The rationale for this choice of $T_\text{w}$ is discussed in Sec.~\ref{subsec:corr}. The duration of the ringdown window is chosen to be three cycles long. 
Since the ringdown GW signal decays rapidly, the reconstructed strain beyond three cycles is dominated by noise and excluded from the analysis. 
This procedure yields an offset of $T_\text{w}-T_\text{p}\approx2 ~t_{M_f}$ from the peak time $T_\text{p}$ of the GW signal, and a constant window duration in the units of remnant mass — corresponding to a longer window for high-mass signals and a shorter window for low-mass signals.




\subsubsection{Moving close to the merger}
\label{subsec:corr}
\begin{figure}[htbp]
    \centering
    \hspace{-0.5cm}
    \includegraphics[width=0.5\textwidth]{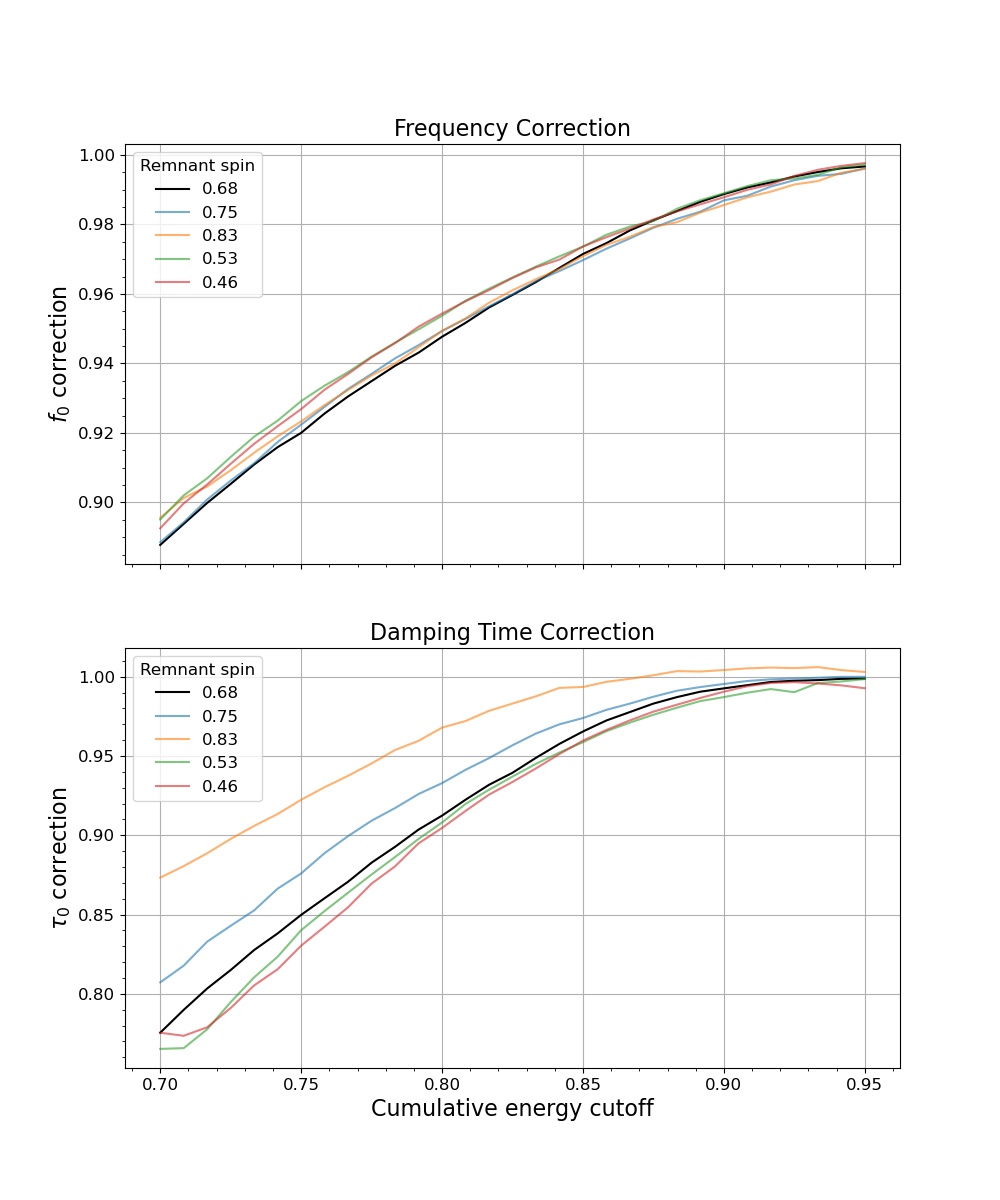}
    \caption{\label{fig:corr_crs} Correction terms for the $(2,2,0)$ mode frequency (top) and damping time (bottom) as a function of the normalized cumulative energy, shown for different values of the remnant spin. Similar correction curves are obtained for $(2,2,1)$ mode as well.
    }
\end{figure}

If the ringdown analysis begins close to the peak amplitude of the signal, the inferred parameters may be biased by residual merger effects, because QNMs describe only the linear perturbations of a black hole and the merger dynamics are inherently nonlinear. Typically, the ringdown analysis begins at 8–10 ${t_{M_f}}$ \cite{LIGOScientific:2021sio,LIGOScientific:2026wpt} after the signal peak time $T_{\text{p}}$ to exclude the merger effects. However, the analysis can be performed closer to the peak time if the reconstruction bias in the ringdown parameters is small and can be corrected using simulated GW signals. We define the following correction terms for the frequency and damping time as a function of the normalized cumulative energy, which determines the start time of the ringdown window,
\begin{align}
    f_0 \text{ correction} &=  \frac{f_0}{f^{\text{GR}}_{220}}, \\  \tau_0 \text{ correction} &=  \frac{\tau^{\text{GR}}_{220}}{\tau_{0}}.
\end{align}    
Here, $f^{\text{GR}}_{220}$ and $\tau^{\text{GR}}_{220}$ denote the expected frequency and damping time, respectively, of the $(2,2,0)$ ringdown mode. Analogous correction terms are defined for the $(2,2,1)$ mode. The correction terms are evaluated by fitting the ringdown model described in Sec. \ref{subsec:ring_model} to simulated GW signals 
generated using the \texttt{SEOBNRv4HM} waveform approximant~\cite{Ramos-Buades:2021adz,Cotesta:2020qhw}. 
As the normalized cumulative energy increases, the window moves further from the peak time, thereby reducing the bias introduced by the merger. Consequently, the correction terms approach unity, as illustrated in Fig.~\ref{fig:corr_crs}. 

The correction terms are independent of the system's total mass and depend weakly on other source properties, such as mass ratio and component spins; they may also vary slightly across waveform approximants. Nonzero spin projections along the orbital axis alter the merger dynamics and the resulting remnant spin. This effect is amplified for unequal-mass mergers, causing the remnant spin to deviate from the value of $\sim{0.68}$ expected for non-spinning components. As shown in Fig.~\ref{fig:corr_crs}, the correction curves are practically independent of the component spins, except for aligned-spin mergers that produce a rapidly rotating remnant.
The correction curves corresponding to the zero-spin components are adopted for the ringdown analysis.
As a consequence, higher remnant spins might be underestimated. However, this bias is significantly smaller than the associated damping-time uncertainties and yields a more conservative estimate of the remnant spin.  For $e(t)=0.82$, the correction for both frequency and damping time is  $4\%$ qnd $7\%$ respectively.  The small magnitude of the correction ensures that its uncertainties have a negligible impact on the inferred ringdown parameters. Consequently, the correction terms can be treated as universal and applied to all BBH events considered in this analysis.




\section{Simulations}
\label{sim}


Simulations were used to assess potential reconstruction biases and systematic uncertainties arising from cWB waveform reconstruction, as well as to study the impact of non-Gaussian detector noise on the estimated ringdown parameters. The simulated signals were generated with \texttt{SEOBNRv4HM} waveform approximant~\cite{Ramos-Buades:2021adz,Cotesta:2020qhw} with the mass ratio $q=1$, non-spinning components, and zero inclination angle. The sources were distributed uniformly over the sky, with network SNR ranging from 12 to 50. The simulated signals were injected into O4 detector data and reconstructed using cWB. Ringdown analyses were then performed on the reconstructed waveforms, and the recovered parameters were compared to the injected values to characterize the resulting biases and systematic uncertainties.

\subsection{Reconstruction bias}
\label{subsec:rec_bias}

The cWB reconstruction is subject to systematic uncertainties associated with the selection of the TF pixels. Imposing a threshold to select significant pixels helps extract a gravitational wave from the strain data, yielding a denoised estimation of the signal. 
However, this selection process can also result in incomplete recovery of the signal power and distortion of the reconstructed waveform, introducing additional systematic biases and uncertainties in the inferred ringdown parameters. The ringdown frequency is particularly sensitive to this effect, as the inclusion or exclusion of the TF pixels in the vicinity of the ringdown signal directly impacts its estimation. The ringdown damping time is less affected since the wavelets used in the TF transform are much longer than a typical damping time of BBH remnants. Therefore, inclusion or exclusion of a few TF pixels has negligible effect on the reconstructed damping time. 

The selection of the TF pixels produces two kinds of systematic biases.
First, the loss of TF pixels near the ringdown signal leads to a systematic underestimation of the reconstructed ringdown frequency for low-SNR signals. We apply an SNR-dependent correction, dividing the reconstructed ringdown frequencies by $(1 - e^{-\text{SNR}/5.9})$ to compensate for this bias.

Second, the early merger signal suffers a greater loss of power for low-SNR signals and high-mass signals, which changes the shape of the cumulative energy curve and shifts the ringdown window away from the peak of the signal. To mitigate this effect for the low-SNR signals, we modify the equation for $T_{\text{w}}$ with the SNR-dependent correction
\begin{equation}
\label{eq:cumulative_cutoff}
    e(T_{\text{w}})=\left(1 - \frac{N}{\text{3  SNR}^2}\right) 0.82,
\end{equation}
where $N$ is the number of independent TF data samples comprising the reconstructed event. This correction ensures that, on average, the ringdown window offset from the peak amplitude is independent on the signal SNR. A similar loss of power for high-mass events, where the low-frequency merger signal is obscured by the seismic detector noise, moves the window from the nominal $2~t_{M_f}$ offset by an additional  time shift
\begin{equation}
  \Delta{t({M_{\text{t}}}}) \approx 0.028\cdot{M_{\text{t}}} - 1.06,
\end{equation}
as a function of the detector-frame total mass $M_\text{t}$. The value of $M_\text{t}$ is estimated prior to the ringdown analysis by using the rapid parameter estimation with the cWB machine-learning regression model~\cite{PhysRevD.104.082003}. The corrected window start time is given than by $T_{\text{w}}-\Delta{t({M_{\text{t}}}})$. The correction becomes negligible for systems with  $M_{\text{t}} < 100~M_\odot$. 
Consequently, the uncertainty in the estimated total mass has little impact on the window correction.

\subsection{Systematic Errors}
\label{subsec:sys_err}
\begin{table*}[ht]
\renewcommand{\arraystretch}{1.4}
\begin{tabular}{lrrrrrccc}
\hline
\textbf{Events} & $\delta f_{220}$ & $\delta \tau_{220}$ &
$f_{220}$ (Hz) & $\tau_{220}$ (ms) &
$M_f/M_\odot$ & $\chi_f$ & $\epsilon$ & SNR  \\
\hline
GW150914 & $0.06_{-0.08}^{+0.08}$ & $-0.20_{-0.17}^{+0.31}$ & $267.9_{-18.3}^{+19.4}$ & $3.2_{-0.6}^{+1.3}$ & $56.3_{-9.2}^{+14.2}$ & $0.52_{-0.39}^{+0.27}$ & $0.55_{-0.43}^{+0.29}$ & 24  \\ 
GW170104 & $-0.08_{-0.11}^{+0.12}$ & $0.42_{-0.54}^{+1.00}$ & $267.0_{-29.0}^{+31.4}$ & $5.0_{-1.8}^{+3.3}$ & $73.2_{-18.5}^{+18.7}$ & $0.83_{-0.39}^{+0.12}$ & $0.37_{-0.33}^{+0.50}$ & 13  \\ 
GW190408\_181802 & $-0.01_{-0.13}^{+0.12}$ & $0.02_{-0.29}^{+0.66}$ & $311.5_{-39.0}^{+32.0}$ & $3.3_{-0.9}^{+2.1}$ & $54.3_{-11.8}^{+16.2}$ & $0.67_{-0.46}^{+0.23}$ & $0.52_{-0.46}^{+0.41}$ & 15  \\ 
GW190503\_185404 & $0.08_{-0.16}^{+0.17}$ & $-0.06_{-0.32}^{+0.56}$ & $210.6_{-25.1}^{+26.8}$ & $4.8_{-1.3}^{+2.7}$ & $79.4_{-17.3}^{+24.1}$ & $0.66_{-0.37}^{+0.22}$ & $0.45_{-0.39}^{+0.40}$ & 11  \\ 
GW190513\_205428 & $0.04_{-0.18}^{+0.19}$ & $-0.12_{-0.33}^{+0.58}$ & $249.6_{-31.3}^{+37.2}$ & $3.9_{-1.0}^{+2.4}$ & $65.6_{-14.3}^{+21.7}$ & $0.63_{-0.37}^{+0.26}$ & $0.39_{-0.35}^{+0.48}$ & 13  \\ 
GW190519\_153544 & $0.05_{-0.11}^{+0.11}$ & $-0.18_{-0.26}^{+0.37}$ & $136.7_{-13.1}^{+12.0}$ & $7.5_{-1.7}^{+3.1}$ & $124.2_{-21.3}^{+28.4}$ & $0.69_{-0.34}^{+0.18}$ & $0.76_{-0.35}^{+0.20}$ & 14  \\ 
GW190521\_030229 & $0.02_{-0.11}^{+0.11}$ & $0.55_{-0.59}^{+0.73}$ & $70.2_{-6.3}^{+6.7}$ & $21.7_{-6.4}^{+9.3}$ & $299.1_{-53.1}^{+45.5}$ & $0.88_{-0.15}^{+0.07}$ & $0.53_{-0.45}^{+0.39}$ & 14  \\ 
GW190521\_074359 & $-0.09_{-0.07}^{+0.06}$ & $0.24_{-0.25}^{+0.27}$ & $181.5_{-10.4}^{+9.0}$ & $6.5_{-1.2}^{+1.4}$ & $101.9_{-11.8}^{+10.8}$ & $0.78_{-0.18}^{+0.09}$ & $0.95_{-0.14}^{+0.04}$ & 24  \\ 
GW190602\_175927 & $-0.01_{-0.14}^{+0.16}$ & $-0.05_{-0.35}^{+0.42}$ & $103.1_{-10.4}^{+15.5}$ & $9.7_{-2.7}^{+3.9}$ & $161.3_{-37.6}^{+39.2}$ & $0.66_{-0.39}^{+0.19}$ & $0.77_{-0.41}^{+0.20}$ & 12  \\ 
GW190706\_222641 & $0.07_{-0.14}^{+0.14}$ & $-0.06_{-0.33}^{+0.41}$ & $116.4_{-13.2}^{+10.9}$ & $10.4_{-2.6}^{+4.1}$ & $160.4_{-26.8}^{+28.9}$ & $0.79_{-0.26}^{+0.12}$ & $0.87_{-0.27}^{+0.11}$ & 13 \\ 
GW190828\_063405 & $0.02_{-0.12}^{+0.09}$ & $-0.06_{-0.28}^{+0.46}$ & $245.2_{-25.2}^{+19.4}$ & $4.4_{-1.2}^{+2.1}$ & $71.1_{-14.2}^{+15.3}$ & $0.71_{-0.47}^{+0.19}$ & $0.88_{-0.40}^{+0.11}$ & 16  \\ 
GW190915\_235702 & $0.08_{-0.14}^{+0.14}$ & $-0.02_{-0.31}^{+0.54}$ & $256.5_{-32.5}^{+28.9}$ & $4.3_{-1.2}^{+2.5}$ & $68.9_{-14.9}^{+18.6}$ & $0.73_{-0.43}^{+0.18}$ & $0.55_{-0.49}^{+0.40}$ & 12  \\ 
GW191109\_010717 & $-0.00_{-0.10}^{+0.10}$ & $0.31_{-0.37}^{+0.50}$ & $120.3_{-9.6}^{+9.6}$ & $10.2_{-2.1}^{+3.7}$ & $156.9_{-21.1}^{+27.0}$ & $0.80_{-0.17}^{+0.10}$ & $0.61_{-0.30}^{+0.21}$ & 17  \\ 
GW191222\_033537 & $0.02_{-0.15}^{+0.13}$ & $0.17_{-0.36}^{+0.53}$ & $152.7_{-18.4}^{+13.9}$ & $7.9_{-2.3}^{+3.5}$ & $122.4_{-22.9}^{+25.2}$ & $0.79_{-0.34}^{+0.12}$ & $0.81_{-0.43}^{+0.17}$ & 12  \\ 
GW200224\_222234 & $-0.02_{-0.09}^{+0.09}$ & $0.14_{-0.27}^{+0.45}$ & $192.7_{-16.2}^{+14.6}$ & $6.4_{-1.4}^{+2.5}$ & $98.3_{-13.7}^{+17.5}$ & $0.80_{-0.19}^{+0.11}$ & $0.66_{-0.41}^{+0.27}$ & 18  \\ 
GW200311\_115853 & $-0.03_{-0.09}^{+0.09}$ & $0.24_{-0.30}^{+0.45}$ & $230.2_{-20.5}^{+16.9}$ & $5.4_{-1.3}^{+1.9}$ & $82.7_{-12.1}^{+13.0}$ & $0.80_{-0.22}^{+0.10}$ & $0.89_{-0.30}^{+0.10}$ & 16  \\ 
\hline
\end{tabular}
\caption{Results of the ringdown analysis for selected GW events from GWTC-3. The deviations from the IMR predictions for the $(2,2,0)$ mode frequency $(\delta f_{220})$ and damping time $(\delta \tau_{220})$ are reported, together with the ringdown parameters in the detector frame, including the $(2,2,0)$ mode frequency, $f_{220}$, damping time, $\tau_{220}$, detector frame remnant mass, $M_f$, remnant spin, $\chi_f$, and the relative amplitude, $\epsilon$ of the $(2,2,1)$ QNM.  The quoted uncertainties correspond to $90\%$ credible intervals and include all identified systematic uncertainties. The total error is obtained by adding the statistical and systematic errors in quadrature, as described in Section \ref{subsec:sys_err}.
}
\label{O3_res}
\end{table*}
The estimation of the reference time is prone to uncertainties due to noise in the reconstructed signal, which in turn introduces systematic errors in the measured ringdown parameters. We estimate the corresponding uncertainties by using injections of simulated signals into real data. Specifically, for each injection, we measure the difference $T_\text{w}-T_\text{p}$. 
Deviations of $T_\text{w}-T_\text{p}$ from its mean characterize the jitter in the reference time, which in turn introduces a systematic uncertainty in the reconstructed ringdown parameters.
Using simulations, we estimate the 90\% confidence intervals for the jitters $\sigma_{e}$  and $\sigma_p$ of the reference time defined by the cumulative energy and the envelope peak time, respectively:
\begin{equation}
    \sigma_{e} \sim \frac{129}{\text{SNR}}~ t_{M_f}, \quad \sigma_p \sim\frac{389}{\text{SNR}}~t_{M_f}.
\end{equation}

The jitter $\sigma_p$ is substantially larger. For instance, at a SNR of $25$, we find, $\sigma_e \sim 5~ t_{M_{f}} $ and $\sigma_p \sim 15~t_{M_{f}}$. Furthermore, the distribution $T_{\text{w}}-T_{\text{p}}$ associated with the envelope peak time is non-Gaussian and exhibits a long tail. These results indicate that the reference time based on the cumulative energy is less susceptible to noise fluctuations than the envelope peak time.

The $90\%$ confidence intervals for the systematic uncertainties in the frequency, $\sigma_f$, and the damping time, $\sigma_{\tau}$, induced by the reference-time jitter are given by,
\begin{equation}
\label{f_tau_sys}
\sigma_f \sim \frac{120}{\text{SNR}}\%,~~~
\sigma_{\tau} \sim \frac{300}{\text{SNR}}\%.
\end{equation}
The total uncertainty in the ringdown parameters quoted in this paper is obtained by adding the statistical and systematic uncertainties in quadrature. 

To validate the estimated uncertainties, simulations are used to evaluate the coverage of a credible interval,  defined as the proportion of trials in which the true value is contained within the interval. Ideally, the coverage should be equal to the nominal credibility level of the interval. 

\begin{figure}[h]
    \centering  
    \hspace{-0.6cm}
    \includegraphics[width=0.5\textwidth]{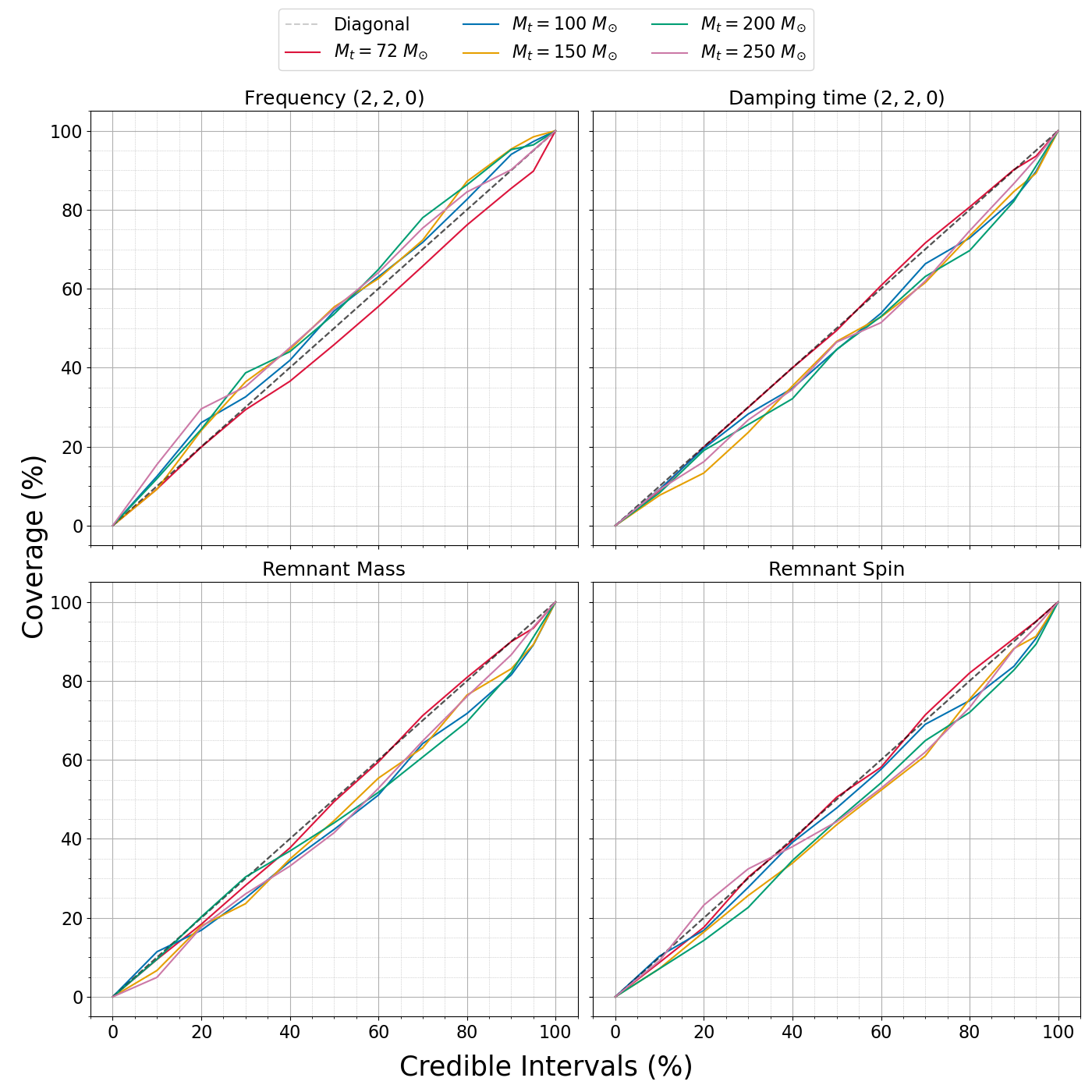}
    \caption{\label{fig:1d_cov} Coverage of the marginalized posteriors for the $(2,2,0)$ mode - frequency, damping time, remnant mass, and spin - across the mass range 72--250 $M_t$. }
\end{figure}
Fig.~\ref{fig:1d_cov} shows the coverage of the marginalized posteriors for the $(2,2,0)$ mode frequency, damping time, remnant mass and spin across a wide range of total masses $M_t$. All estimated quantities show good agreement with the expected coverage across the entire mass range

\section{Results}

In this section, we present the results of the ringdown analysis for a subset of events from the GWTC-3~\cite{KAGRA:2021vkt}. We consider events with $\mathrm{SNR}>12$ and total mass $M>50M_\odot$, as inferred from the IMR analysis. These selection criteria ensure sufficient signal strength in the ringdown phase. The measured ringdown parameters are compared with the corresponding IMR ringdown parameters, which are derived within GR from the inferred source properties and therefore represent the GR predictions. We define the fractional deviations from GR as follows,
\begin{equation}
\begin{aligned}
\delta f_{220} &= \frac{f_{220} - f^{\mathrm{IMR}}_{220}}{f^{\mathrm{IMR}}_{220}}, \\
\delta \tau_{220} &= \frac{\tau_{220} - \tau^{\mathrm{IMR}}_{220}}{\tau^{\mathrm{IMR}}_{220}} ,
\end{aligned}
\end{equation}
where $f_{220}$ and $\tau_{220}$ are the $(2,2,0)$ QNM frequency and damping times, respectively, obtained from the ringdown analysis. While $f^{\mathrm{IMR}}_{220}$ and $\tau^{\mathrm{IMR}}_{220}$ are the corresponding GR-predicted values calculated using the IMR analysis. The $90\%$ credible intervals of these deviations are then used to test GR by assessing their consistency with zero. Table~\ref{O3_res} shows the results of the ringdown analysis for events from the GWTC-3. The uncertainties in  Table \ref{O3_res} correspond to $90\%$ credible intervals, including all systematic errors. 

All 16 events are consistent with the IMR predictions, with only one event, \textnormal{GW190521\_074359} exhibiting small deviation in frequency beyond the $90\%$ credible intervals. Given the number of events considered, these deviations are consistent with statistical fluctuations. Most events exhibit a significant contribution from the $(2,2,1)$ mode, characterized by the relative amplitude $\epsilon$, which must be taken into account to avoid bias in the reconstructed ringdown parameters. Combining the measurements from all 16 events yields tighter constraints on the deviations from the IMR predictions. The joint deviations with $90\%$ credible intervals are 
\begin{equation}
\delta f_{220} = 
-0.005_{-0.028}^{+0.028} \quad 
\delta\tau_{220} =
0.032_{-0.090}^{+0.108}.
\end{equation}
The reconstructed remnant spins for all the events considered are broadly consistent with the values expected for binary black holes (BBHs) formed through isolated stellar evolution. In particular, such systems are predicted to have low component spins, resulting in a remnant spin close to 0.68. As shown in Table~\ref{O3_res}, the reconstructed spins for all events are consistent with this value, with the exception of \textnormal{GW190521\_030229}, which shows evidence for a higher spin. This might indicate a possible origin through dynamical formation channels, which is further supported by the fact that the primary component mass of this system lies within the pair-instability mass gap~\cite{PhysRevD.104.082003,PhysRevLett.125.101102,PhysRevD.103.082002}.

In addition to individual measurements, the joint remnant spin provides a more stringent test of consistency with the predictions of stellar evolution. However, the remnant spin depends nonlinearly on the ringdown frequency and damping time, resulting in asymmetric uncertainties that become increasingly restrictive at higher spin values. Consequently, constructing a joint spin posterior directly from the individual spin measurements can introduce a bias toward larger spins.
To avoid this bias, the joint posterior is obtained for the product $f_{220}\cdot\tau_{220}$, which depends only on the remnant spin. This resulting joint distribution is then mapped to the corresponding posterior for the remnant spin. Combining all events in this manner, we obtain the average remnant spin, with $90\%$ credible intervals,
\begin{equation}
    \chi_f = 0.72_{-0.06}^{+0.06}.
\end{equation}
This result shows consistency with the canonical value of $0.68$ expected for black holes formed through stellar evolution. 

\section{Conclusion}

We have presented \texttt{RingCWB}, a ringdown analysis framework designed to improve the precision of remnant black hole measurements while minimizing the impact of detector noise and waveform-model systematics. The method combines denoised signal reconstruction with cWB, a robust and model-independent definition of the ringdown window based on the cumulative signal energy, and a dedicated study of systematic uncertainties using simulated signals injected into real detector data.

Simulations show that the cumulative-energy reference time is substantially less sensitive to noise fluctuations than the commonly used signal envelope peak time, allowing the ringdown analysis to begin at approximately $2~t_{M_f}$ after the merger. This leads to improved precision in the measurement of the QNM parameters. Systematic uncertainties associated with the reference-time jitter and the cWB reconstruction were quantified and incorporated into the final parameter estimates. Coverage studies demonstrate that the resulting uncertainties provide reliable confidence intervals for the inferred ringdown parameters.

We applied this method to 16 BBH events from the GWTC-3 with $\mathrm{SNR}>12$ and total mass $M>50~M_\odot$. For all considered events, the measured frequency and damping time of the dominant $(2,2,0)$ mode are consistent with the predictions of general relativity. Combining the measurements from all events yields stringent bounds on the deviations, which show consistency with zero.

The reconstructed remnant spins for all events are consistent with the expectations from stellar evolution, with the exception of \textnormal{GW190521\_030229}, which exhibits a higher remnant spin.  This is consistent with alternative formation channel, such as dynamical formation, discussed in the literature~\cite{PhysRevD.104.082003,PhysRevLett.125.101102,PhysRevD.103.082002}. In addition to individual measurements, the combined remnant spin estimate is consistent with the value 0.68 expected for black holes formed through stellar evolution. 

This method yields tighter constraints on the QNM frequency and damping time than previous ringdown analysis of the GWTC-4 events~\cite{LIGOScientific:2026wpt}. The improved precision is enabled by the noise reduction provided by the cWB reconstruction and by the ability to probe the remnant properties closer to the merger. With the increasing sensitivity of GW detectors and the growing population of detected binary black hole mergers, the approach presented here provides a promising framework for precision ringdown measurements and increasingly stringent tests of general relativity.

\section*{Acknowledgment}
This material is based upon work supported by NSF's LIGO Laboratory which is a major facility fully funded by the National Science Foundation. We gratefully acknowledge the support of LIGO and Virgo for the provision of computational resources, especially LIGO Laboratory, which is supported by the NSF Grant No. PHY $0757058$ and PHY $0823459$. This work was supported by the NSF grant No. PHY-2409372. M.S. acknowledges Polish National Science Centre Grants No. UMO-2023/49/B/ST9/02777 and No. UMO-2024/03/1/ST9/00005, and the Polish National Agency for Academic Exchange within Polish Returns Programme Grant No. BPN/PPO/2023/1/00019. This research has made use of data, software, and/or web tools obtained from the Gravitational Wave Open Science Center, a service of LIGO Laboratory, the LIGO Scientific Collaboration, and the Virgo Collaboration. This manuscript has the LIGO preprint No. P2600322. We acknowledge the valuable feedback from Tanmaya Mishra, N. V. Krishnendu and Gayathri.V.   

\bibliographystyle{apsrev}
\bibliography{main}
\end{document}